\input epsf
\documentclass[12pt]{article}

\newcommand{\bsgluon}     {\mbox{$b\! \rightarrow \! s \> gluon  $}}
\newcommand{\bsg}     {\mbox{$b\! \rightarrow \! sg$}}
\newcommand{\kpm}     {\mbox{$K^{\pm}$}}
\newcommand{\pipm}     {\mbox{$\pi^{\pm}$}}

\newcommand{\btoc}    {\mbox{$b\! \rightarrow \! c$}}
\newcommand{\btos}    {\mbox{$b\! \rightarrow \! s$}}
\newcommand{\btoctos}    {\mbox{$b\! \rightarrow \! c \! \rightarrow  \! s$}}

\newcommand{\brdc}    {\mbox{${\cal B}(B\rightarrow D\bar DX)$}}
\newcommand{\brbsg}  {\mbox{${\cal B}(b\! \rightarrow \! sg)$}}
\newcommand{\bplus}     {\mbox{$B^+$}}
\newcommand{\bzero}     {\mbox{$B^0$}}
\newcommand{\bzerod}     {\mbox{$B^0_d$}}
\newcommand{\wus}      {\mbox{$W\! \rightarrow \! u \! \bar s$}}
\newcommand{\bpsix}    {\mbox{$B\! \rightarrow \! J/\Psi X$}}
\newcommand{\psimumu}  {\mbox{$J/\Psi \! \rightarrow \! \mu^+\mu^-$}}
\newcommand{\bsl}  {\mbox{${\cal B}_{SL}$}}
\newcommand{\tableline} {\hline}

\begin{document}        

\pagestyle{empty}
\renewcommand{\thefootnote}{\fnsymbol{footnote}}
 
\begin{flushright}
{\small
SLAC--PUB--8076\\
March 1999\\}
\end{flushright}
 
\vspace{.8cm}
 
\begin{center}
{\bf\large   
B Decay Studies at SLD\footnote{Work supported by
Department of Energy contract  DE--AC03--76SF00515.}}
 
\vspace{1cm}

M. R. Convery\\
Stanford Linear Accelerator Center, Stanford University,
Stanford, CA  94309\\
\vskip .2in
Representing the SLD Collaboration$^{**}$
\medskip
\end{center}
 
\vfill
 
\begin{center}
{\bf\large   
Abstract }
\end{center}

We present three preliminary results from SLD on $B$ decays: 
an inclusive search for the process \bsgluon,
a measurement of the 
branching ratio for the process $B\! \rightarrow \! D\bar DX$, 
and measurements of the charged and neutral $B$ lifetimes. 
All three measurements make use of the excellent 
vertexing efficiency and resolution
of the CCD Vertex Detector
and the first two make use of the excellent
particle identification capability of the 
Cherenkov Ring Imaging Detector.
The \bsg\ analysis searches for an enhancement of high 
momentum charged kaons produced in $B$ decays. 
Within the context of a 
simple, Jetset-inspired model of \bsg, 
a limit of \brbsg $< 7.6$\% is obtained.
The \brdc\ analysis reconstructs two secondary vertices and uses
identified charged kaons to determine which of these came from charm 
decays. 
The result of the analysis is 
\brdc = $(16.2 \pm 1.9 \pm 4.2)$\%.
The results of the lifetime analysis are:
\hbox{$\tau_{B^+} = 1.686 \pm 0.025 \pm 0.042\;$ ps,}
\hbox{$\tau_{B^0} = 1.589 \pm 0.026 \pm 0.055\;$ ps}
 and
\hbox{$\tau_{B^+}/\tau_{B^0} = 1.061 \pm^{0.031}_{0.029} \pm 0.027$.}

\vfill
 
\begin{center} 
{\it Presented at the American Physical Society (APS) Meeting of 
the Division of Particles and Fields (DPF 99), 5-9 January 1999, 
University of California, Los Angeles} 
\end{center}

\newpage

 
 
%
\pagestyle{plain}

\section{Introduction}               

\label{sec:intro}

Detailed studies of $B$-hadron decays can provide important
tests of the Standard Model.
In fact, it has been suggested that the existence of 
several ``$B$-decay puzzles''
may be pointing the way to physics beyond the Standard Model
\cite{newphysics}.
The most serious of these puzzles is the 
the low measured value compared to theoretical expectations 
of the the $B$ semi-leptonic branching ratio:\cite{drell_lp97}
\begin{eqnarray}
{
{\cal B}_{SL} = 
{\Gamma_{semi-leptonic}
\over
\Gamma_{semi-leptonic} +
\Gamma_{hadronic} +
\Gamma_{leptonic}
}
}
\end{eqnarray}
Theoretical expectations of ${\cal B}_{SL}$ typically have a lower
limit of 12.5\%
\cite{bigi}.
As of Summer '97, however, the experimental measurements were much
lower:
the world average of measurements done at the $\Upsilon(4S)$ was
$0.1018 \pm 0.0040$ and at the $Z^0$ was $0.1095 \pm 0.0032$
\cite{drell_lp97}.

Since $\Gamma_{semi-leptonic}$ is well understood theoretically
and $\Gamma_{leptonic}$ is very small, efforts to explain the
discrepancy have focused on $\Gamma_{hadronic}$, which can
be broken into three parts (assuming that Cabibbo suppressed rates
are small):
\begin{eqnarray}
 \Gamma_{hadronic} \approx
 \Gamma(b \! \rightarrow \! c \bar u d) + 
 \Gamma(b \! \rightarrow \! c \bar c s) +
 \Gamma(b \! \rightarrow \! s g)
\end{eqnarray}
Reducing \bsl\ to the experimentally measured value would require
enhancing one or more of these components significantly
above the expected value.
In this paper, we describe measurements that bear on each of these
three components.

\subsection{SLD Capabilities and Data Set}
The SLD experiment collects $Z^0$ decay data from $e^+e^-$ 
collisions at the SLAC Linear Collider with a center of mass
energy of 91.28 GeV. 
A full description of the SLD detector may be found in \cite{sldref}.
SLD is well-suited for doing precision measurements
of inclusive $B$-decays due to several unique characteristics.
The SLC interaction point is small and stable and its position
is known with an uncertainty of 5 $\mu$m transverse to the beam 
direction.
This precise IP is complemented by SLD's high precision
CCD vertex detector, VXD3 \cite{vxd3}.
For high momentum tracks the impact parameter resolution is
$\sigma(r\phi) = 11 \mu$m and $\sigma(rz) = 22 \mu$m. 
Multiple scattering adds a momentum-dependent contribution of 
$33 \mu{\rm m}/(p \sin^{3/2}(\theta))$, where $p$ is the momentum
expressed in GeV/c and $\theta$ is the track polar angle.
Note that the above describes the performance of the upgraded
vertex detector (VXD3) installed prior to the start of the 1996 
run. 
For the performance of the vertex detector used before 1996 (VXD2),
see Reference \cite{sldref}.
Another important capability of SLD is the excellent particle
identification provided by the Cherenkov Ring Imaging Detector
\cite{crid}.
\kpm's in the barrel region with momentum between 1 and 20 GeV/c 
are identified with an efficiency of 50\% and a $\pipm$ 
misidentification probability of 2\%.
Data used in these analyses was taken between 1993 and 1998. 
However, not all data has yet been used in all analyses.
Table \ref{tab:dataset} lists the different running periods
used for each analysis.

\begin{table}

\begin{center}
\begin{tabular}{lll|ccc}
Running Period  &  Number of Z's  & Vertex Detector & \bsg & \brdc &
$\tau_{\bplus}$,$\tau_{\bzero}$ \\
\hline
1993-95   & 150K       & VXD2     & Yes & No & Yes \\
1996-98   & 250K       & VXD3     & Yes & Yes & Yes \\
Spring 98 & 150K       & VXD3     & Yes & Yes & No 
\end{tabular}
\end{center}
\caption{\label{tab:dataset} The three running periods used in
the analyses are listed along with the vertex detector in use
for that period and whether the period was used in each of the
three analyses}
\end{table}

\subsection{Analysis Techniques}
\label{sec:anal}
All three of the analyses described in this paper make use of 
an inclusive $B$ reconstruction method.
This method was originally developed for the SLD $R_b$ measurement
and is described in detail in \cite{sldrb}.
Briefly, the procedure is as follows.
Well measured tracks with vertex detector hits are selected.
In each hemisphere, secondary vertices are formed from 
these ``quality'' tracks using a topological vertexing 
technique \cite{dj}.
The most significant of these vertices that is 
significantly displaced from the IP is chosen as the
``seed''.
The $B$ flight direction is then defined by the line
joining the primary vertex and this secondary vertex.
Additional tracks are attached to the vertex if they
satisfy the following criteria.

\begin{itemize}
\item{the 3D closest approach to the flight direction is $<$ 1mm}
\item{the distance along the flight direction to this point, L, is $>$0.5 mm}
\item{the $L/D >$ 0.25, where D is the secondary vertex decay
distance. }
\end{itemize}

\begin{figure}[h]      
\centerline{\epsfysize 2.3 truein \epsfbox{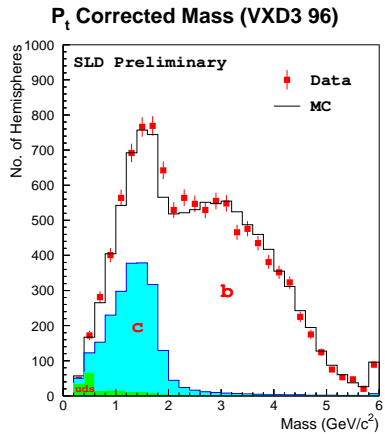}}   
\caption[]{
\label{fig:rbmass96}
\small $P_T$ corrected mass for '96 data and Monte Carlo.
Cutting at $M_{P_T} > 2 {\rm GeV/c^2}$ gives a very pure and efficient $b$ tag.}
\end{figure}

These tracks, seed plus attached, are called ``$B$-tracks'' and
a $b$-tag is formed by calculating their invariant mass
assuming that each has $m_{\pipm}$.
A correction is applied to account for neutrals and missing tracks
which is based on the total transverse momentum of the vertex tracks
to the $B$ flight direction. 
Figure \ref{fig:rbmass96} shows a histogram of this ``$P_T$ corrected mass'',
$M_{P_T}$.
Finally a cut on $M_{P_T}$ is applied at 2 ${\rm GeV/c^2}$. 
In Monte Carlo studies, this $b$ selection method
had an efficiency of 35\% and a purity of 98\% in '93-'95 and
an efficiency of 50\% and a purity of 98\% in '96-'98.

In addition to the $M_{P_T}$ cut, the set of ``$B$-tracks'' can also 
be used to study the structure of the $B$ decay.
This is done by fitting all $B$-tracks to a single vertex and 
calculating the fit probability.
Due to the finite charm lifetime, $B$ decays with open charm will
tend to have lower fit probability than those that are ``charmless''.
Decays with fit probability greater than 0.05 are called 
``1-Vertex'' and the remaining decays are called ``2-Vertex''.
Table \ref{tab:vtx} shows that, indeed, decays without open charm
are more likely to be in the 1-Vertex sample.

\begin{table}
\begin{center}
\begin{tabular}{lc}
B Decay Mode   & 1-Vertex Fraction\\
\tableline
Single Charm   & 0.33 $\pm$ 0.01 \\
Double Charm   & 0.16 $\pm$ 0.02 \\
Charmonium + X & 0.76 $\pm$ 0.03 \\
\bsg\ Model    & 0.73 $\pm$ 0.01
\end{tabular}
\end{center}
\caption{\label{tab:vtx} Monte Carlo estimates of the fraction of $B$ decays of 
different types that satisfy the 1-Vertex cut.}
\end{table}

A check is also performed on data in \bpsix\ data where \psimumu.
In these events, the 1-Vertex fraction is found to be 
0.733 $\pm$ 0.094, confirming the efficiency estimated from the 
Monte Carlo.

\section{Inclusive Search for \bsg}
In the Standard Model, \bsg\ occurs through gluonic penguins and 
is expected to have a total branching ratio of approximately 1\%
\cite{lenz}.
However, if the branching ratio were enhanced up to $\approx$ 
10 \% by some non-Standard Model mechanism, it would nicely explain
the ``puzzles'' described in section \ref{sec:intro}
\cite{newphysics}.
Experimentally, it has been difficult to exclude even such a
large \brbsg\ due to the lack of a clean signature for these
decays.

The SLD analysis \cite{ich_paper}
searches for such a large enhancement by examining
the high $p_t$ part of the \kpm\ spectrum, where $p_t$ is the 
momentum transverse to the $B$ flight direction.
Naively, one would expect \kpm's produced by \bsg\ to have a
stiffer spectrum than those produced from standard $B$ decays
since they come from a direct \btos\ transition rather
than cascade \btoctos\ transitions.
In a simple, {\sc jetset} \cite{jetset} inspired model \cite{alex}
it has been verified that this is indeed the case.
The model, however, is quite sensitive to the choice of tuning 
of {\sc jetset}.
Table \ref{tab:bsgmod} shows the number of high-$p_t$ \kpm's 
expected for different choices of tuning.
``{\sc delphi} Tuning'' will be used for the rest of the analysis, but
the tuning sensitivity means that limits on \bsg\ can be set
only within the context of a particular tuning choice.

\begin{table}
\begin{center}
\begin{tabular}{l|c}
Tuning Choice & \kpm,$p_t > 1.8$ GeV/c per B\\
\tableline
{\sc jetset} default     & $6.9 \times 10^{-3}$  \\
{\sc delphi} tuning      & $10.6 \times 10^{-3}$ \\
No Parton showers  & $13.8 \times 10^{-3}$ \\
\tableline
\tableline
\btoc & $2.4 \times 10^{-3}$ \\
\end{tabular}
\caption{\label{tab:bsgmod} The number of high $p_t$ \kpm's
expected from \bsg\ for a number of different choices of 
{\sc jetset} tunings.}
\end{center}
\end{table}

\begin{figure}[h]      
\centerline{\epsfysize 4.0 truein \epsfbox{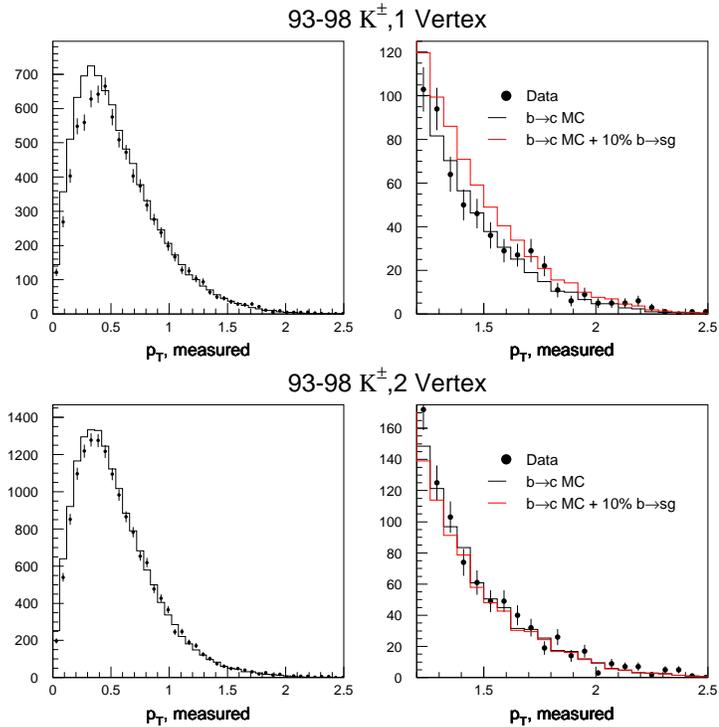}}   
\caption[]{
\label{fig:spec}
\small The $p_t$ spectrum of identified \kpm's in the 1- and
2-Vertex samples. The data is compared to \btoc\ MC background and
to \btoc\ background plus a 10\% \bsg\ contribution.}
\end{figure}

In addition to the $M_{P_T}$ cut described above,
the 1-vertex cut is also used. 
Since \bsg\ events are charmless, they are expected to be
mostly one 1-vertex.
Using the 1-vertex cut thus provides background rejection
of standard \btoc\ decays as well as providing a 2-vertex
sample, which should not have much \bsg\ in it and 
can be used to check the background calculation.
The \bsg\ signal should therefore show up as an enhancement 
in high $p_t$ \kpm's in the 1-Vertex sample.
Only the high $p_t$ part of the spectrum is used both because
it has high signal to background and also because the 
background in that region is well understood, leading to 
a small systematic error.
Figure \ref{fig:spec} shows the spectrum of \kpm's observed
in both the 1- and 2-Vertex samples for a sample of
50623 inclusively reconstructed $B$'s.
Table \ref{tab:spec} shows the number of events expected from
\btoc\ background and the number observed in the 1-Vertex 
sample and for all events.
The data is well described by the \btoc\ Monte Carlo with only
a small excess of high $p_t$ \kpm's observed.

\begin{table}
\begin{center}
\begin{tabular}{l|c|c|c|c}
\kpm,$p_t > 1.8$ GeV/c& \multicolumn{2}{c|}{1 vertex}&\multicolumn{2}{c}{All}\\
\tableline
& Raw & Normalized & Raw & Normalized\\
\tableline
Data       &
53.0&
$1.46\times 10^{-3}$&
150.0&
$4.18\times 10^{-3}$\\

\btoc\ MC& 
48.2&
$1.29\times 10^{-3}$&
132.3 &
$3.54\times 10^{-3}$ \\
\tableline
Difference & 
4.8
       $\pm$ 
7.3 &
$( 1.7 \pm 2.6 ) \times 10^{ -4 }$ &
17.7
       $\pm$ 
12.2 & 
$( 6.4 \pm 4.4 ) \times 10^{ -4 }$ \\
\tableline
 BR(\bsg)=10\% &
31.5 &
$1.13\times 10^{-3}$ &
42.2 &
$1.52\times 10^{-3}$ \\
\end{tabular}
\end{center}
\caption{\label{tab:spec} The number \kpm's with $p_t>1.8$ GeV/c observed
in data and expected from the \btoc\ Monte Carlo with and without
the 1-Vertex cut. Normalized numbers are per $B$ decay
and are corrected for \kpm\ identification efficiency}
\end{table}

The systematic error on the background rate is calculated by
splitting the background high $p_t$ \kpm's into their component 
sources and assigning an error to each one of these sources.
The largest source of background is from \kpm's from 
$D^0$'s that were produced in $B$-decays.
This source accounts for 23.5 of the expected 48.2 background
events.
Fortunately, the spectrum of such $D^0$'s has been rather well
measured \cite{cleo-dprod} and the SLD $B$-decay model 
has been tuned to match it.
This leads to a rather small systematic error of 4.3 events
from this source.
Other large sources of systematic error include uncertainties
in the branching ratios of Cabibbo suppressed $B$ decays of the
type \wus\ (3.6 events), $p_t$ smearing (4.5 events) 
and \kpm\ identification efficiency (3.3 events).

The preliminary result is that the excess of 1-vertex
\kpm's with $p_t>1.8$ GeV/c is $4.8 \pm 7.3 (stat.) \pm 8.9 (syst.)$ events.
Within the context of the \bsg\ model with {\sc delphi} tuning, 
this corresponds
to ${\cal B}(\bsg) = 0.015 \pm 0.023 \pm 0.028$.
Adding the statistical and systematic errors in quadrature,
this gives ${\cal B}(\bsg) <0.076$ at 90\% confidence.

\section{Measurement of \brdc}
Measurement of the $B$ ``double-charm'' branching ratio (\brdc)
can help to resolve the puzzles described section \ref{sec:intro}.
The SLD measurement of \brdc\ uses the same inclusive reconstruction 
method described above to select a set of $B$-tracks.
The 2-Vertex cut is then applied -
single vertex fit probability less than 0.05.
The $B$-tracks in these decays are then split into two vertices
using a $\chi^2$ minimization procedure.
The upstream vertex is called the ``$B$''-vertex and the downstream
vertex is called the ``$D$''-vertex.
For single charm $B$ decays, the assignment of tracks to these
vertices will tend to be correct - the $B$-vertex will contain 
mostly tracks coming directly from the $B$ decay and the $D$-vertex
will contain mostly tracks coming from a subsequent $D$ decay.
However, for double charm $B$ decays, the $B$-vertex will contain
a mix of $B$ and $D$ tracks.
Since \kpm's come primarily from $D$-decays, if a \kpm\ is found
in the $B$-vertex, it is a good indication that the decay was in
fact a double charm decay.

To extract \brdc\ from the data, the ratio $R_K$ is formed:
\begin{eqnarray}
{
R_K = N_{K,B}/N_{K,D}
}
\end{eqnarray}
where $N_{K,B}$ is the number of \kpm's observed in B-vertices and
$N_{K,D}$ is number observed in D-vertices.
\brdc may then be solved for by comparing $R_K$ observed in data
to $R_K$ found in Monte Carlo samples of
pure double-charm and non-double-charm $B$ decays.
Table \ref{tab:brdc} shows the numbers that go into this calculation.

\begin{table}
\begin{center}
\begin{tabular}{lrrrrl}
& $N_{K^-,B}$ & 
  $N_{K^+,B}$ &
  $N_{K^-,D}$ & 
  $N_{K^+,D}$ &
  $N_{K,B} / N_{K,D}$ \\ 
\tableline
Data  &
1193 & 
871  &
1830 & 
1171 & 
0.688$\pm$ 0.020 \\

M.C. {$D\bar D$} &
3068 & 
2303 &
2363 &
1865 &
1.270$\pm$ 0.026 \\

M.C. ``Not-$D\bar D$'' & 
5301  & 
3903  &
11387 &
 6432 &
0.517$\pm$ 0.007
\end{tabular}
\caption{\label{tab:brdc} Number of \kpm's in the $B$- and $D$-vertices
for data and for Monte Carlo.}
\end{center}
\end{table}

Two corrections are applied to the data:
background from light quark events is subtracted
\hbox{($\Delta\brdc=$ -0.9\%)} and the difference in
efficiency of the 2-Vertex cut between one-charm and two-charm
decays is corrected for \hbox{($\Delta\brdc$=-5.8\%).}

The largest detector related systematic error comes from 
misassignment of tracks between $B$- and $D$-vertices.
This effect is calibrated in the data using an initial
state tag to identify the expected charge of leptons
coming directly from the $B$ decay.
``Right'' sign leptons should then be found in the $B$-vertex
and ``wrong'' sign leptons should be found in the $D$-vertex.
The efficiency for correct track assignment can then be 
extracted from the ratio of correct to incorrect assignment.
Similarly, $B \rightarrow D^{*+}X$ decays, with exclusive
$D^{*+}$ decays are used to identify a set of tracks of
definite origin, which are also used to measure the
track misassignment efficiency.
The statistical error of the track assignment efficiency
is then used to calculate the systematic error due to this 
source, which is found to be 2.2\%.

The largest physics systematics are related to 
$B$-decay modelling and to 
${\cal B}(D\! \rightarrow \! K^{\pm}X)$.
These sources lead to uncertainties of 2.1\% and 1.8\%
respectively.
The preliminary result is then 
\brdc = \hbox{$(16.2  \pm 1.9 (stat.) \pm 4.2 (syst.)) \%$}

\section{Measurement of $\tau_{B^+}$, $\tau_{B^0}$ and $\tau_{B^+}/\tau_{B^0}$}
Measurement of exclusive $B$ lifetimes provides important information
about $B$-hadron decay dynamics.
Deviations from the naive spectator model are expected to be small and
the ratio $\tau_{B^+}/\tau_{B^0}$ is expected to differ
from unity by only about 10\%
\cite{blife}.
Large deviations from unity could, for example, indicate a larger
than expected value for 
$\Gamma(b \! \rightarrow \! c \bar u d)$ \cite{honscheid}.

\begin{figure}      
\centerline{\epsfysize 2.5 truein \epsfbox{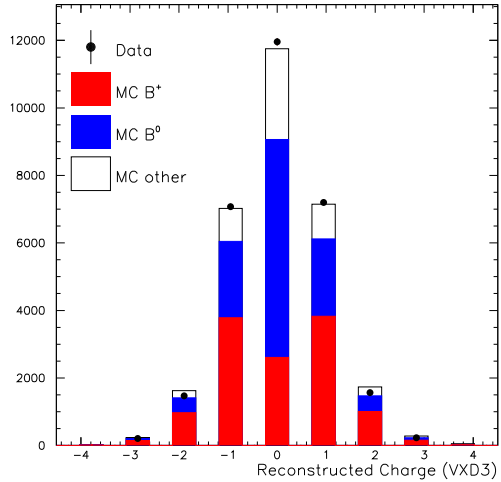}}   
\caption[]{
\label{fig:qvtx9798}
\small Reconstructed charge, $Q$, for '97-'98 data and Monte Carlo.}
\end{figure}

The SLD analysis \cite{sldblife} uses the same inclusive reconstruction
method as described in section \ref{sec:anal}.
In order to improve the charge purity, tracks which
failed the initial quality cuts, but which still likely originated
from the $B$-decay are also included.
The total charge of the $B$-tracks, $Q$ is then calculated and the
reconstructed $B$'s are split into a neutral sample ($Q=0$) and 
a charged sample (Q=$\pm$1,2,3).
Figure \ref{fig:qvtx9798} shows the distribution of $Q$ for data
and Monte Carlo.
Monte Carlo studies show that the ratio between $B^+$ and
$B^0_d$ decays in the charged sample is 1.55 (1.72) for
VXD2 (VXD3).
Similarly, the ratio between $B^0_d$ and $B^+$ in the neutral
sample is 1.96 (2.24) for VXD2 (VXD3).
\footnote{Charge conjugation is implied throughout.}

Since the precision of the measurement depends heavily on the
charge reconstruction purity, several methods are used to
enhance it.
In Monte Carlo studies, it was found that the charge reconstruction
purity depended on the reconstructed $M_{P_T}$, since decays
that are missing some tracks tend to have lower $M_{P_T}$.
Therefore, to enhance the charge purity, events are weighted 
as a function of $M_{P_T}$. 
For charged decays, the polarized forward-backward asymmetry can
be used to tag the $b$ or $\bar b$ flavor of the hemisphere.
The opposite hemisphere jet charge also provides similar information.
If the charge of the decay agrees with that expected from
these tags, the decay is weighted more heavily.
Conversely, if the charge disagrees, the decay is de-weighted.

\begin{figure}      
\centerline{\epsfysize 2.0 truein \epsfbox{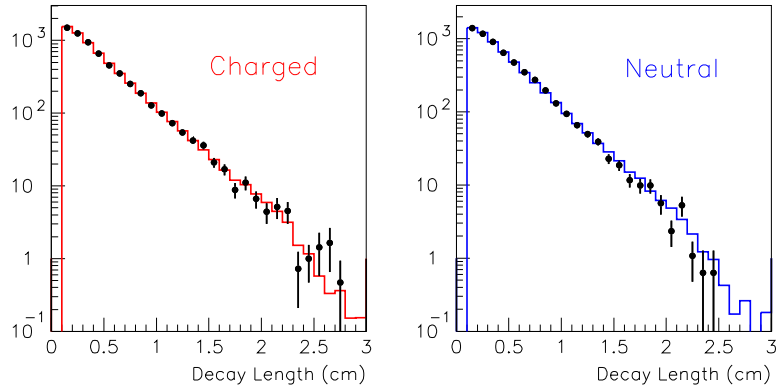}}
\caption[]{
\label{fig:declen}
\small Decay length distributions for charged and neutral samples
in '97-'98 data.}
\end{figure}

The  \bplus\ and \bzerod\ lifetimes are then extracted with
a simultaneous binned maximum likelihood fit to the decay 
length distributions for charged and neutral samples.
Figure \ref{fig:declen} shows these decay length distributions
for data taken in '97-'98.

For the measurements of $\tau_{B^+}$ and $\tau_{B^0}$, the
dominant systematic error is related to $b$ fragmentation.
This is studied by varying both the mean fragmentation energy
$<x_E>$ and the shape of the $x_E$ distribution.
The systematic uncertainty from this source is found to 
be 0.036 ps for both $\tau_{B^+}$ and $\tau_{B^0}$.
For the $\tau_{B^+}/\tau_{B^0}$ measurement, the uncertainty
due to fragmentation largely cancels out, since the two
hadrons are assumed to have the same fragmentation function.
For this measurement, then, the largest systematics are related
to the fraction of $b$-baryons produced 
($\Delta(\tau_{B^+}/\tau_{B^0}) = 0.013$) 
and \brdc\ ($\Delta(\tau_{B^+}/\tau_{B^0})= 0.011$).

The combined '93-'98 preliminary results are then
\begin{eqnarray*}
\tau_{B^+} & = & 1.686 \pm 0.025 \pm 0.042 \; ps, \\
\tau_{B^0} & = & 1.589 \pm 0.026 \pm 0.055 \; ps, \\
\tau_{B^+}/\tau_{B^0} & = & 1.061 \pm^{0.031}_{0.029} \pm 0.027
\end{eqnarray*}
These measurements are the most statistically precise to date and confirm
the expectation that the $B^+$ and $B^0_d$ lifetimes are nearly equal.

\section{Conclusion}
We have presented preliminary results on three analyses of
$B$ hadron decays.
Within the context of the \bsg\ model described in 
\cite{alex} we find ${\cal B}(\bsg) < 0.076$, at 90\% confidence.
We have measured \brdc = \hbox{$(16.2 \pm 1.9 \pm 4.2)$ \%.}
And, we have measured the lifetimes $\tau_{B^+}$, $\tau_{B^0}$ with
the best statistical precision currently available.

Work is continuing on each of the three analyses.
The lifetime analysis will benefit from the addition of 
150,000 more $Z^0$ decays from the Spring '98 running.
All three analyses will benefit from a re-reconstruction of the
data that will incorporate significant tracking improvements.
And finally, more SLD running would significantly improve the
errors of each of the three analyses.


\section*{$^{**}$List of Authors} 


%
%
%
\begin{center}
\def\iADEL{$^{(1)}$}
\def\iAOMORI{$^{(2)}$}
\def\iBOLO{$^{(3)}$}
\def\iBRI{$^{(4)}$}
\def\iBRUN{$^{(5)}$}
\def\iBU{$^{(6)}$}
\def\iCINC{$^{(7)}$}
\def\iCOLO{$^{(8)}$}
\def\iCOLU{$^{(9)}$}
\def\iCSU{$^{(10)}$}
\def\iFERR{$^{(11)}$}
\def\iFRAS{$^{(12)}$}
\def\iILLI{$^{(13)}$}
\def\iJHU{$^{(14)}$}
\def\iLBL{$^{(15)}$}
\def\iLTU{$^{(16)}$}
\def\iMASS{$^{(17)}$}
\def\iMISSI{$^{(18)}$}
\def\iMIT{$^{(19)}$}
\def\iMOSCOW{$^{(20)}$}
\def\iNAGO{$^{(21)}$}
\def\iOREG{$^{(22)}$}
\def\iOXF{$^{(23)}$}
\def\iPADO{$^{(24)}$}
\def\iPERU{$^{(25)}$}
\def\iPISA{$^{(26)}$}
\def\iRAL{$^{(27)}$}
\def\iRUTG{$^{(28)}$}
\def\iSLAC{$^{(29)}$}
\def\iSOGA{$^{(30)}$}
\def\iSOONG{$^{(31)}$}
\def\iTENN{$^{(32)}$}
\def\iTOHO{$^{(33)}$}
\def\iUCSB{$^{(34)}$}
\def\iUCSC{$^{(35)}$}
\def\iUVIC{$^{(36)}$}
\def\iVAND{$^{(37)}$}
\def\iWASH{$^{(38)}$}
\def\iWISC{$^{(39)}$}
\def\iYALE{$^{(40)}$}

  \baselineskip=.75\baselineskip  
\mbox{Kenji  Abe\unskip,\iNAGO}
\mbox{Koya Abe\unskip,\iTOHO}
\mbox{T. Abe\unskip,\iSLAC}
\mbox{I. Adam\unskip,\iSLAC}
\mbox{T.  Akagi\unskip,\iSLAC}
\mbox{N.J. Allen\unskip,\iBRUN}
\mbox{W.W. Ash\unskip,\iSLAC}
\mbox{D. Aston\unskip,\iSLAC}
\mbox{K.G. Baird\unskip,\iMASS}
\mbox{C. Baltay\unskip,\iYALE}
\mbox{H.R. Band\unskip,\iWISC}
\mbox{M.B. Barakat\unskip,\iLTU}
\mbox{O. Bardon\unskip,\iMIT}
\mbox{T.L. Barklow\unskip,\iSLAC}
\mbox{G.L. Bashindzhagyan\unskip,\iMOSCOW}
\mbox{J.M. Bauer\unskip,\iMISSI}
\mbox{G. Bellodi\unskip,\iOXF}
\mbox{R. Ben-David\unskip,\iYALE}
\mbox{A.C. Benvenuti\unskip,\iBOLO}
\mbox{G.M. Bilei\unskip,\iPERU}
\mbox{D. Bisello\unskip,\iPADO}
\mbox{G. Blaylock\unskip,\iMASS}
\mbox{J.R. Bogart\unskip,\iSLAC}
\mbox{G.R. Bower\unskip,\iSLAC}
\mbox{J.E. Brau\unskip,\iOREG}
\mbox{M. Breidenbach\unskip,\iSLAC}
\mbox{W.M. Bugg\unskip,\iTENN}
\mbox{D. Burke\unskip,\iSLAC}
\mbox{T.H. Burnett\unskip,\iWASH}
\mbox{P.N. Burrows\unskip,\iOXF}
\mbox{A. Calcaterra\unskip,\iFRAS}
\mbox{D. Calloway\unskip,\iSLAC}
\mbox{B. Camanzi\unskip,\iFERR}
\mbox{M. Carpinelli\unskip,\iPISA}
\mbox{R. Cassell\unskip,\iSLAC}
\mbox{R. Castaldi\unskip,\iPISA}
\mbox{A. Castro\unskip,\iPADO}
\mbox{M. Cavalli-Sforza\unskip,\iUCSC}
\mbox{A. Chou\unskip,\iSLAC}
\mbox{E. Church\unskip,\iWASH}
\mbox{H.O. Cohn\unskip,\iTENN}
\mbox{J.A. Coller\unskip,\iBU}
\mbox{M.R. Convery\unskip,\iSLAC}
\mbox{V. Cook\unskip,\iWASH}
\mbox{R.F. Cowan\unskip,\iMIT}
\mbox{D.G. Coyne\unskip,\iUCSC}
\mbox{G. Crawford\unskip,\iSLAC}
\mbox{C.J.S. Damerell\unskip,\iRAL}
\mbox{M.N. Danielson\unskip,\iCOLO}
\mbox{M. Daoudi\unskip,\iSLAC}
\mbox{N. de Groot\unskip,\iBRI}
\mbox{R. Dell'Orso\unskip,\iPERU}
\mbox{P.J. Dervan\unskip,\iBRUN}
\mbox{R. de Sangro\unskip,\iFRAS}
\mbox{M. Dima\unskip,\iCSU}
\mbox{A. D'Oliveira\unskip,\iCINC}
\mbox{D.N. Dong\unskip,\iMIT}
\mbox{M. Doser\unskip,\iSLAC}
\mbox{R. Dubois\unskip,\iSLAC}
\mbox{B.I. Eisenstein\unskip,\iILLI}
\mbox{V. Eschenburg\unskip,\iMISSI}
\mbox{E. Etzion\unskip,\iWISC}
\mbox{S. Fahey\unskip,\iCOLO}
\mbox{D. Falciai\unskip,\iFRAS}
\mbox{C. Fan\unskip,\iCOLO}
\mbox{J.P. Fernandez\unskip,\iUCSC}
\mbox{M.J. Fero\unskip,\iMIT}
\mbox{K. Flood\unskip,\iMASS}
\mbox{R. Frey\unskip,\iOREG}
\mbox{J. Gifford\unskip,\iUVIC}
\mbox{T. Gillman\unskip,\iRAL}
\mbox{G. Gladding\unskip,\iILLI}
\mbox{S. Gonzalez\unskip,\iMIT}
\mbox{E.R. Goodman\unskip,\iCOLO}
\mbox{E.L. Hart\unskip,\iTENN}
\mbox{J.L. Harton\unskip,\iCSU}
\mbox{A. Hasan\unskip,\iBRUN}
\mbox{K. Hasuko\unskip,\iTOHO}
\mbox{S.J. Hedges\unskip,\iBU}
\mbox{S.S. Hertzbach\unskip,\iMASS}
\mbox{M.D. Hildreth\unskip,\iSLAC}
\mbox{J. Huber\unskip,\iOREG}
\mbox{M.E. Huffer\unskip,\iSLAC}
\mbox{E.W. Hughes\unskip,\iSLAC}
\mbox{X. Huynh\unskip,\iSLAC}
\mbox{H. Hwang\unskip,\iOREG}
\mbox{M. Iwasaki\unskip,\iOREG}
\mbox{D.J. Jackson\unskip,\iRAL}
\mbox{P. Jacques\unskip,\iRUTG}
\mbox{J.A. Jaros\unskip,\iSLAC}
\mbox{Z.Y. Jiang\unskip,\iSLAC}
\mbox{A.S. Johnson\unskip,\iSLAC}
\mbox{J.R. Johnson\unskip,\iWISC}
\mbox{R.A. Johnson\unskip,\iCINC}
\mbox{T. Junk\unskip,\iSLAC}
\mbox{R. Kajikawa\unskip,\iNAGO}
\mbox{M. Kalelkar\unskip,\iRUTG}
\mbox{Y. Kamyshkov\unskip,\iTENN}
\mbox{H.J. Kang\unskip,\iRUTG}
\mbox{I. Karliner\unskip,\iILLI}
\mbox{H. Kawahara\unskip,\iSLAC}
\mbox{Y.D. Kim\unskip,\iSOGA}
\mbox{M.E. King\unskip,\iSLAC}
\mbox{R. King\unskip,\iSLAC}
\mbox{R.R. Kofler\unskip,\iMASS}
\mbox{N.M. Krishna\unskip,\iCOLO}
\mbox{R.S. Kroeger\unskip,\iMISSI}
\mbox{M. Langston\unskip,\iOREG}
\mbox{A. Lath\unskip,\iMIT}
\mbox{D.W.G. Leith\unskip,\iSLAC}
\mbox{V. Lia\unskip,\iMIT}
\mbox{C.Lin\unskip,\iMASS}
\mbox{M.X. Liu\unskip,\iYALE}
\mbox{X. Liu\unskip,\iUCSC}
\mbox{M. Loreti\unskip,\iPADO}
\mbox{A. Lu\unskip,\iUCSB}
\mbox{H.L. Lynch\unskip,\iSLAC}
\mbox{J. Ma\unskip,\iWASH}
\mbox{G. Mancinelli\unskip,\iRUTG}
\mbox{S. Manly\unskip,\iYALE}
\mbox{G. Mantovani\unskip,\iPERU}
\mbox{T.W. Markiewicz\unskip,\iSLAC}
\mbox{T. Maruyama\unskip,\iSLAC}
\mbox{H. Masuda\unskip,\iSLAC}
\mbox{E. Mazzucato\unskip,\iFERR}
\mbox{A.K. McKemey\unskip,\iBRUN}
\mbox{B.T. Meadows\unskip,\iCINC}
\mbox{G. Menegatti\unskip,\iFERR}
\mbox{R. Messner\unskip,\iSLAC}
\mbox{P.M. Mockett\unskip,\iWASH}
\mbox{K.C. Moffeit\unskip,\iSLAC}
\mbox{T.B. Moore\unskip,\iYALE}
\mbox{M.Morii\unskip,\iSLAC}
\mbox{D. Muller\unskip,\iSLAC}
\mbox{V. Murzin\unskip,\iMOSCOW}
\mbox{T. Nagamine\unskip,\iTOHO}
\mbox{S. Narita\unskip,\iTOHO}
\mbox{U. Nauenberg\unskip,\iCOLO}
\mbox{H. Neal\unskip,\iSLAC}
\mbox{M. Nussbaum\unskip,\iCINC}
\mbox{N. Oishi\unskip,\iNAGO}
\mbox{D. Onoprienko\unskip,\iTENN}
\mbox{L.S. Osborne\unskip,\iMIT}
\mbox{R.S. Panvini\unskip,\iVAND}
\mbox{C.H. Park\unskip,\iSOONG}
\mbox{T.J. Pavel\unskip,\iSLAC}
\mbox{I. Peruzzi\unskip,\iFRAS}
\mbox{M. Piccolo\unskip,\iFRAS}
\mbox{L. Piemontese\unskip,\iFERR}
\mbox{K.T. Pitts\unskip,\iOREG}
\mbox{R.J. Plano\unskip,\iRUTG}
\mbox{R. Prepost\unskip,\iWISC}
\mbox{C.Y. Prescott\unskip,\iSLAC}
\mbox{G.D. Punkar\unskip,\iSLAC}
\mbox{J. Quigley\unskip,\iMIT}
\mbox{B.N. Ratcliff\unskip,\iSLAC}
\mbox{T.W. Reeves\unskip,\iVAND}
\mbox{J. Reidy\unskip,\iMISSI}
\mbox{P.L. Reinertsen\unskip,\iUCSC}
\mbox{P.E. Rensing\unskip,\iSLAC}
\mbox{L.S. Rochester\unskip,\iSLAC}
\mbox{P.C. Rowson\unskip,\iCOLU}
\mbox{J.J. Russell\unskip,\iSLAC}
\mbox{O.H. Saxton\unskip,\iSLAC}
\mbox{T. Schalk\unskip,\iUCSC}
\mbox{R.H. Schindler\unskip,\iSLAC}
\mbox{B.A. Schumm\unskip,\iUCSC}
\mbox{J. Schwiening\unskip,\iSLAC}
\mbox{S. Sen\unskip,\iYALE}
\mbox{V.V. Serbo\unskip,\iSLAC}
\mbox{M.H. Shaevitz\unskip,\iCOLU}
\mbox{J.T. Shank\unskip,\iBU}
\mbox{G. Shapiro\unskip,\iLBL}
\mbox{D.J. Sherden\unskip,\iSLAC}
\mbox{K.D. Shmakov\unskip,\iTENN}
\mbox{C. Simopoulos\unskip,\iSLAC}
\mbox{N.B. Sinev\unskip,\iOREG}
\mbox{S.R. Smith\unskip,\iSLAC}
\mbox{M.B. Smy\unskip,\iCSU}
\mbox{J.A. Snyder\unskip,\iYALE}
\mbox{H. Staengle\unskip,\iCSU}
\mbox{A. Stahl\unskip,\iSLAC}
\mbox{P. Stamer\unskip,\iRUTG}
\mbox{H. Steiner\unskip,\iLBL}
\mbox{R. Steiner\unskip,\iADEL}
\mbox{M.G. Strauss\unskip,\iMASS}
\mbox{D. Su\unskip,\iSLAC}
\mbox{F. Suekane\unskip,\iTOHO}
\mbox{A. Sugiyama\unskip,\iNAGO}
\mbox{S. Suzuki\unskip,\iNAGO}
\mbox{M. Swartz\unskip,\iJHU}
\mbox{A. Szumilo\unskip,\iWASH}
\mbox{T. Takahashi\unskip,\iSLAC}
\mbox{F.E. Taylor\unskip,\iMIT}
\mbox{J. Thom\unskip,\iSLAC}
\mbox{E. Torrence\unskip,\iMIT}
\mbox{N.K. Toumbas\unskip,\iSLAC}
\mbox{T. Usher\unskip,\iSLAC}
\mbox{C. Vannini\unskip,\iPISA}
\mbox{J. Va'vra\unskip,\iSLAC}
\mbox{E. Vella\unskip,\iSLAC}
\mbox{J.P. Venuti\unskip,\iVAND}
\mbox{R. Verdier\unskip,\iMIT}
\mbox{P.G. Verdini\unskip,\iPISA}
\mbox{D.L. Wagner\unskip,\iCOLO}
\mbox{S.R. Wagner\unskip,\iSLAC}
\mbox{A.P. Waite\unskip,\iSLAC}
\mbox{S. Walston\unskip,\iOREG}
\mbox{J. Wang\unskip,\iSLAC}
\mbox{S.J. Watts\unskip,\iBRUN}
\mbox{A.W. Weidemann\unskip,\iTENN}
\mbox{E. R. Weiss\unskip,\iWASH}
\mbox{J.S. Whitaker\unskip,\iBU}
\mbox{S.L. White\unskip,\iTENN}
\mbox{F.J. Wickens\unskip,\iRAL}
\mbox{B. Williams\unskip,\iCOLO}
\mbox{D.C. Williams\unskip,\iMIT}
\mbox{S.H. Williams\unskip,\iSLAC}
\mbox{S. Willocq\unskip,\iMASS}
\mbox{R.J. Wilson\unskip,\iCSU}
\mbox{W.J. Wisniewski\unskip,\iSLAC}
\mbox{J. L. Wittlin\unskip,\iMASS}
\mbox{M. Woods\unskip,\iSLAC}
\mbox{G.B. Word\unskip,\iVAND}
\mbox{T.R. Wright\unskip,\iWISC}
\mbox{J. Wyss\unskip,\iPADO}
\mbox{R.K. Yamamoto\unskip,\iMIT}
\mbox{J.M. Yamartino\unskip,\iMIT}
\mbox{X. Yang\unskip,\iOREG}
\mbox{J. Yashima\unskip,\iTOHO}
\mbox{S.J. Yellin\unskip,\iUCSB}
\mbox{C.C. Young\unskip,\iSLAC}
\mbox{H. Yuta\unskip,\iAOMORI}
\mbox{G. Zapalac\unskip,\iWISC}
\mbox{R.W. Zdarko\unskip,\iSLAC}
\mbox{J. Zhou\unskip.\iOREG}

\it
  \vskip \baselineskip                   
  \centerline{(The SLD Collaboration)}   
  \vskip \baselineskip        
  \baselineskip=.75\baselineskip   
\iADEL
  Adelphi University, Garden City, New York 11530, \break
\iAOMORI
  Aomori University, Aomori , 030 Japan, \break
\iBOLO
  INFN Sezione di Bologna, I-40126, Bologna, Italy, \break
\iBRI
  University of Bristol, Bristol, U.K., \break
\iBRUN
  Brunel University, Uxbridge, Middlesex, UB8 3PH United Kingdom, \break
\iBU
  Boston University, Boston, Massachusetts 02215, \break
\iCINC
  University of Cincinnati, Cincinnati, Ohio 45221, \break
\iCOLO
  University of Colorado, Boulder, Colorado 80309, \break
\iCOLU
  Columbia University, New York, New York 10533, \break
\iCSU
  Colorado State University, Ft. Collins, Colorado 80523, \break
\iFERR
  INFN Sezione di Ferrara and Universita di Ferrara, I-44100 Ferrara, Italy, \break
\iFRAS
  INFN Lab. Nazionali di Frascati, I-00044 Frascati, Italy, \break
\iILLI
  University of Illinois, Urbana, Illinois 61801, \break
\iJHU
  Johns Hopkins University,  Baltimore, Maryland 21218-2686, \break
\iLBL
  Lawrence Berkeley Laboratory, University of California, Berkeley, California 94720, \break
\iLTU
  Louisiana Technical University, Ruston,Louisiana 71272, \break
\iMASS
  University of Massachusetts, Amherst, Massachusetts 01003, \break
\iMISSI
  University of Mississippi, University, Mississippi 38677, \break
\iMIT
  Massachusetts Institute of Technology, Cambridge, Massachusetts 02139, \break
\iMOSCOW
  Institute of Nuclear Physics, Moscow State University, 119899, Moscow Russia, \break
\iNAGO
  Nagoya University, Chikusa-ku, Nagoya, 464 Japan, \break
\iOREG
  University of Oregon, Eugene, Oregon 97403, \break
\iOXF
  Oxford University, Oxford, OX1 3RH, United Kingdom, \break
\iPADO
  INFN Sezione di Padova and Universita di Padova I-35100, Padova, Italy, \break
\iPERU
  INFN Sezione di Perugia and Universita di Perugia, I-06100 Perugia, Italy, \break
\iPISA
  INFN Sezione di Pisa and Universita di Pisa, I-56010 Pisa, Italy, \break
\iRAL
  Rutherford Appleton Laboratory, Chilton, Didcot, Oxon OX11 0QX United Kingdom, \break
\iRUTG
  Rutgers University, Piscataway, New Jersey 08855, \break
\iSLAC
  Stanford Linear Accelerator Center, Stanford University, Stanford, California 94309, \break
\iSOGA
  Sogang University, Seoul, Korea, \break
\iSOONG
  Soongsil University, Seoul, Korea 156-743, \break
\iTENN
  University of Tennessee, Knoxville, Tennessee 37996, \break
\iTOHO
  Tohoku University, Sendai 980, Japan, \break
\iUCSB
  University of California at Santa Barbara, Santa Barbara, California 93106, \break
\iUCSC
  University of California at Santa Cruz, Santa Cruz, California 95064, \break
\iUVIC
  University of Victoria, Victoria, British Columbia, Canada V8W 3P6, \break
\iVAND
  Vanderbilt University, Nashville,Tennessee 37235, \break
\iWASH
  University of Washington, Seattle, Washington 98105, \break
\iWISC
  University of Wisconsin, Madison,Wisconsin 53706, \break
\iYALE
  Yale University, New Haven, Connecticut 06511. \break

\rm
%

\end{center}

\end{document}